\documentclass[conference]{IEEEtran}
\IEEEoverridecommandlockouts
\usepackage{amsmath,amsfonts}
\usepackage{algorithmic}
\usepackage{algorithm}
\usepackage{array}
\usepackage[caption=false,font=scriptsize]{subfig}
\usepackage{textcomp}
\usepackage{stfloats}
\usepackage{url}
\usepackage{verbatim}
\usepackage{graphicx}
\usepackage{textcmds}
\usepackage{cite}
\usepackage{hhline}
\usepackage{bm}
\usepackage{color}
\usepackage{longtable}	
\usepackage{multicol}
\usepackage{float}
\usepackage{multirow}
\usepackage [c1]{optidef}
\usepackage{xspace}
\usepackage{cite}
\usepackage{amsmath,amssymb,amsfonts}
\usepackage{algorithmic}
\usepackage{tikz}
\usetikzlibrary{shapes.geometric, arrows}
\usepackage{textcomp}
\usepackage{diagbox}
\usepackage{xcolor}
\usepackage{graphicx}
\usepackage{nicefrac}
\usepackage{textcmds}
\usepackage{bm}
\usepackage{amsmath,amssymb}
\usepackage{enumerate}
\usepackage{graphicx}
\usepackage{tikz}
\usetikzlibrary{shapes.geometric, arrows}
\usepackage{hhline}
\usepackage{longtable}	
\usepackage{nicefrac}
\usepackage{multicol}
\usepackage{float}
\usepackage{multirow}
\usepackage{epsf,psfrag,amssymb,amsfonts,color,cite,fancybox}
\usepackage{array}
\usepackage[mathscr]{eucal}
\usepackage{algorithmic}
\tikzstyle{startstop} = [rectangle, rounded corners, minimum width=1.2cm, minimum height=0.8cm,text centered, draw=black, fill=white!20]
\tikzstyle{io} = [trapezium, trapezium left angle=70, trapezium right angle=110, minimum width=0cm, minimum height=0cm, text centered, draw=black, fill=red!30]
\tikzstyle{process} = [rectangle, minimum width=0cm, minimum height=0cm, text centered, draw=black, fill=orange!15, rounded corners]
\tikzstyle{decision} = [diamond, minimum width=1.2cm, minimum height=0.8cm, text centered, draw=black, fill=blue!10]
\tikzstyle{arrow} = [thick,->,>=stealth]

\definecolor{geo}{rgb}{0, 0, 0}
\usepackage{textcomp}
\usepackage{diagbox}
\usepackage{xcolor}
\usepackage{soul}
\usepackage{enumitem}
\allowdisplaybreaks

\usepackage{optidef}
\begin{document}

\title{\color{geo}Integrating Optimal EV Charging in the Energy Management of Electric Railway Stations


}


\author{Georgia Pierrou,~\IEEEmembership{Member,~IEEE,} Gabriela Hug,~\IEEEmembership{Senior Member,~IEEE}
\thanks{This work is supported by the ETH Mobility Initiative under MI-GRANT 2020-HS-396. }
\thanks{Georgia Pierrou and Gabriela Hug are with the Power Systems Laboratory, Department of Information Technology and Electrical Engineering, ETH Zurich, Zurich, 8092, Switzerland. (email: gpierrou@ethz.ch, ghug@ethz.ch). 
}

}

\maketitle

\begin{abstract} \color{geo} In this paper, an electric railway Energy Management System (EMS) with integration of an Energy Storage System (ESS), Regenerative Braking Energy (RBE), and renewable generation is proposed to minimize the daily operating costs of the railway station while meeting railway and Electric Vehicle (EV) charging demand. Compared to other railway EMS methods, the proposed approach integrates an optimal EV charging policy at the railway station to avoid high power demand due to charging requirements. Specifically, receding horizon control is leveraged to minimize the daily peak power spent on EV charging. The numerical study on an actual railway station in Chur, Switzerland shows that the proposed method that integrates railway demand and optimal EV charging along with ESS, RBE, and renewable generation
can significantly reduce the average daily operating cost of the railway station over a large number of different scenarios while ensuring that peak load capacity limits are respected.
\color{black}

\end{abstract}
\color{geo}
\begin{IEEEkeywords}
Electric vehicles, energy management, railway systems, receding horizon, regenerative braking energy
\end{IEEEkeywords}
\color{black}

\section{Introduction}

\color{geo}
The consideration of recent advances in transportation electrification with the widespread adoption of electric vehicles (EVs) for private and commercial purposes is crucial for the development of the electric power system infrastructure. In fact, uncontrolled EV charging poses new challenges to power systems, as uncoordinated EV requirements may lead to unwanted phenomena, such as frequent overloading and line congestion. At the same time, there is a growing interest in leveraging the electric railway infrastructure for EV charging, as an opportunity to exploit the parking lots and central location of electric railway stations along with Regenerative Braking Energy (RBE) capabilities, Energy Storage Systems (ESS) and renewable generation at the station level to improve system efficiency and savings \cite{7587629, 9024071}. To this end, novel railway Energy Management Systems (EMS) and smart charging strategies are needed to optimize energy utilization and maintain railway grid stability.

Several solutions have been proposed to address electric railway system operation in the presence of ESS, RBE, and renewable generation in the literature. A multi-period optimal power flow problem formulation to study the potential for energy and economic savings in electric railways with renewable generation and ESS is presented in \cite{7482762}. In \cite{9782535}, a power flow control strategy for railway substations utilizing Model Predictive Control to maximize RBE and photovoltaic (PV) energy utilization and improve power quality is proposed. A railway EMS based on a Mixed Integer Linear Programming (MILP) model incorporating ESS, RBE, PV, and several
pricing schemes is described in \cite{Sengor18}.  

While the energy analysis and management of electric railway stations with any ESS, RBE, and renewable generation have been studied in the previously mentioned works, the integration of EV charging in electric railway operation has attracted less attention. A binary
linear programming method for EV charging in railway station parking lots is presented in \cite{Sarabi19}, which can achieve effective energy utilization and economic benefit. However, other components, such as ESS, are not considered. Moreover, the impact of electricity price variations is not evident, as the price during most of the plug-in intervals is constant. In \cite{Fernandez17}, an operating prototype of the railway system infrastructure including ESS for EV
charging is demonstrated. Nevertheless, renewable generation is not included in the analysis. An integrated rail system and EV parking lot operation with RBE, ESS, and PV generation is proposed in \cite{9745034}. Yet, renewable uncertainty as well as the potential of selling excess power back to the main grid are not included in the formulation. Therefore, a systematic method to coordinate the aforementioned concepts and optimize system operation under varying conditions while ensuring the fulfillment of technical limitations, such as preventing overloading in the combined railway and EV charging requirements, is lacking in the literature. 

In this paper, a novel railway EMS algorithm incorporating ESS, RBE, and renewable generation to serve railway and EV demand and minimize the daily operating costs of the railway station is proposed. The main contributions of the work are the following:

\begin{itemize}
    \item A receding horizon formulation for optimal EV charging is integrated into the railway EMS, with the primary objective of reducing the daily peak charging power that can be crucial when technical constraints, such as line limits for the combined railway and EV demand, arise. 
     \item  The proposed approach considers the uncertainty of input data, such as renewable generation availability and variations in electricity prices in the set of scenarios.
     \item Numerical studies are conducted on an actual railway station in Chur, Switzerland to show that the proposed
EMS method leads to a decrease in the average daily operating cost of the railway station across various scenarios.
\end{itemize}
\newpage
The rest of the paper is organized as follows: Section
II provides the railway energy management mathematical model. Section III reviews the optimal EV charging policy. In Section IV, the proposed EMS algorithm integrating optimal EV charging requirements is presented. Section V validates the proposed algorithm through a comprehensive numerical study. Section VI summarizes the conclusions.

\section{Railway Energy Management Model}

\subsection{Power Balance}
The railway EMS aims to coordinate power exchange from the main grid, available solar generated power, and power that is charged or discharged by ESS to supply internal railway demand and EV charging demand at the train station. Hence, the following power balance constraints should hold: 
\begin{equation}
\label{eq:powerbalance}
P_G^{t,s}+P_{PV}^{t,s}+P_{B^-}^{t,s} = P_{D}^{t,s} + P_{EV}^{t,s} + P_{B^+}^{t,s}+P_{S}^{t,s} \quad  \forall t, s
\end{equation}
where $P_G^{t,s}$ is the power supply from the main grid, $P_{PV}^{t,s}$ the solar generated power, $P_{B^-}^{t,s}$ the ESS discharging power, $P_{D}^{t,s}$ the railway demand, $P_{EV}^{t,s}$ the EV charging demand, $P_{B^+}^{t,s}$ the ESS charging power, and $P_S^{t,s}$ the power sold to the main grid. Superscript $t$ denotes the time step, i.e., $t=1,2,3$ with the time being $t \cdot \Delta t$, $\Delta t$ is the time step size, and superscript $s$ denotes the scenario considered.

\subsection{Power Exchange}
The railway EMS should take into account the fact that power cannot be bought from the main grid or sold back to the grid during the same time interval. Thus, the following power exchange limits should be included in the railway EMS: 
\begin{eqnarray}
\label{eq:gridlimits}
P_{G}^{t,s}&\leq& \bar{P}_{G}u_{G}^{t,s} \quad  \forall t, s \\
\label{eq:gridlimits2}
P_{S}^{t,s}&\leq& \bar{P}_{S}(1-u_{G}^{t,s})  \quad  \forall t, s
\end{eqnarray}
where $\bar{P}_{G}$ is the maximum amount of power that can be bought from the main grid, $u_{G}^{t,s}$ is a binary variable to determine the status of the power exchange, and $\bar{P}_{S}$ is the maximum amount of power that can be sold to the main grid.

\subsection{Solar Generated Power}
In this work, we assume that solar generation is available at the train station where EMS is implemented. Connected solar inverters are typically set up to inject power at unity power factor \cite{WECC}, meaning they only produce active power. Historical data for solar radiation may be used to estimate the solar generated active power as follows \cite{Pierrouu19}:

\begin{equation}
\label{eq:solar_p_val}
P_{PV}^{t,s}={P}_{PV}^{t,s}(\beta^{t,s})=\left\{ \begin{array}{*{35}{l}}
\displaystyle \frac{{{\beta^{t,s}}^{2}}}{{{r}_{c}}{{r}_{std}}}{{P}_{r}} & 0\le\beta^{t,s}<{{r}_{c}}  \\
\displaystyle \frac{\beta^{t,s}}{{{r}_{std}}}{{P}_{r}} & {{r}_{c}}\le \beta^{t,s} < {{r}_{std}}  \\
{{P}_{r}} & \beta^{t,s} \ge {{r}_{std}}  \\
\end{array} \right.
\end{equation}
where $\beta^{t,s}$ is the measured solar radiation, ${{r}_{c}}$ a radiation threshold up to which solar radiation greatly affects solar generation, ${{r}_{std}}$ the solar radiation in the standard environment where radiation increase does not have an impact on solar generation, and ${{P}_{r}}$ is the installed solar capacity at the train station.

\subsection{RBE Modeling}
The railway EMS should aim to use as much as possible of the available power from Regenerative Braking (RB). However, this may not always be possible due to limitations on the ESS capacity. The utilized RB power should be limited to the available RB power using the following constraints:
\begin{eqnarray}
\label{eq:rbelimits}
P_{RBE}^{t,s}&\leq& \bar{P}_{RBE}^{t,s}  \quad  \forall t, s
\end{eqnarray}
where $\bar{P}_{RBE}^{t,s}$ denotes the available RB power and $P_{RBE}^{t,s}$ denotes the RB power that is eventually utilized for charging the ESS.

\subsection{ESS Charging and Discharging Modeling}
In this work, available RB power may be stored in the ESS rather than being wasted \cite{Sengor18, 8260299}. For this purpose, the following constraints are added to the railway EMS regarding the ESS charging, discharging, and state of energy:
\begin{eqnarray}
\label{eq:ESS1}
 P_{RBE}^{t,s} + P_{B^+}^{t,s} &\leq& \bar{P}_{B+}u_{B}^{t,s} \quad  \forall t,s \\
 \label{eq:ESS2}
P_{B^-}^{t,s}&\leq& \bar{P}_{B-}(1-u_{B}^{t,s}) \quad  \forall t,s \\
 \label{eq:ESS0}
P_{B^+}^{t,s}, P_{B^-}^{t,s}&\geq& 0 \quad  \forall t,s \\
\label{eq:ESS3}
SoC_{B}^{t,s} &=& SoC_{B}^{t-1,s} - \epsilon_{B-}SoC_{B}^{t-1,s} \\
\label{eq:ESS4}\hspace{30pt} & & + \hspace{3pt} \eta_{B+}(P_{RBE}^{t,s}+ P_{B^+}^{t,s})\Delta t \nonumber \\ & & - \hspace{3pt} \eta_{B-}P_{B^-}^{t,s}\Delta t \quad  \forall t,s \nonumber\\
\label{eq:ESS5}
SoC_{B}^{t,s} &=& SoC_{B}^{0} \quad  \forall t=t_0\\
\label{eq:ESS6}
SoC_{B}^{t,s}  &\leq& SoC_{B}^{max} \quad  \forall t,s\\
\label{eq:ESS7}
SoC_{B}^{t,s}  &\geq& SoC_{B}^{min} \quad  \forall t,s
\end{eqnarray}
where $SoC_{B}^{t,s}$ is the state of energy of the ESS and $u_{B}^{t,s}$ determines the charging or discharging status and prevents simultaneous charging and discharging, i.e., it is 1 when ESS charges and 0 when ESS discharges. Parameter $\epsilon_{B-}$ is the self-discharge parameter, $\eta_{B+}, \eta_{B-}$ the charging and discharging efficiencies of the ESS, respectively, $SoC_{B}^{0}$ the starting energy level of the ESS, $SoC_{B}^{min}$ the minimum energy limit for ESS discharging, and $SoC_{B}^{max}$ the maximum energy limit for ESS charging.

\subsection{Objective Function}
The objective function of the railway EMS focuses on minimizing the daily operating cost of the train station based on the day-ahead electricity market. It is worth noting that multiple scenarios corresponding to different solar radiation, demand, and price evolutions may be considered, leading to the following objective function:

\begin{equation} 
 \underset {} 
 {\text{minimize}} \quad \sum_s \sum _t \pi_s \hspace{0.025in} (C_{G}^{t,s}P_G^{t,s}-C_{S}^{t,s}P_S^{t,s})\Delta t 
 \label{eq:objfunction}
\end{equation}
where $\pi_s$ is the probability of each scenario, $C_{G}^{t,s}$ the buying electricity price, and $C_{S}^{t,s}$ the selling electricity price.

\color{black}

\color{geo}
\section{EV Charging Policy}
In this paper, a receding horizon approach \cite{Casini21} is incorporated in the railway EMS so that daily peak power consumed for EV charging purposes is minimized. Briefly speaking, assuming that the arrival and departure times of EVs plugged-in at the parking lot of the main train station and the amount of energy to be charged are uncertain, a charging power schedule for each unit satisfying maximum power limits and customer requirements in terms of charged energy is decided. 

\subsection{EV Arrival Framework}
The set of plugged-in EVs to be charged at time $t$ is defined as:
\begin{equation}
\label{eq:pluggedinEVs}
\Omega_{EV}^t = \{ v: t_v^a \leq t < t_v^d, SoC_v^t<SoC_v^f \}
\end{equation}
where $t_v^a$ is the arrival time, $t_v^d$ the departure time, $SoC_v^t$ the current state of charge of vehicle $v$, and $SoC_v^f$ the demanded state of charge of vehicle $v$. Note that the energy level for each plugged-in vehicle is linked to its charging power $P_v^t$ with an efficiency of $\eta_v$ as follows:
\begin{equation}
\label{eq:EVSoC}
SoC_v^{t+1}=SoC_v^{t}+ \eta_v P_v^{t}\Delta t
\end{equation}

Depending on the energy requirements, the power needed to charge the plugged-in EVs at time $t$ at a constant rate (but limited by the  maximum rate $\bar{P}_v$ for each vehicle) can be calculated as:
\begin{equation}
\label{eq:powerneed}
\tilde{\lambda}=\sum_{v \in \Omega_{EV}^t} \text{min} \{ \bar{P}_v, \frac{SoC_v^f-SoC_v^{t}}{ \eta_v \Delta t} \}
\end{equation}

The fulfillment time for each plugged-in vehicle assuming a constant nominal charging rate $P_v^0$ is:
\begin{equation}
\label{eq:fulfilmenttime}
t_v^f = t_v^a + \frac{SoC_v^f-SoC_v^{t_v^a}}{ \eta_v P_v^0 \Delta t}
\end{equation}

However, considering the uncertainty of the departure times, it is possible that a customer may leave earlier or later than the fulfillment time. Therefore, a satisfaction threshold should ensure that during departure, EVs should have either the demanded state of charge or at least a state of charge corresponding to nominal charging. Hence, the customer satisfaction threshold is defined as follows:
\begin{equation}
\label{eq:satisfthreshold}
\theta_v^t =  \text{min} \{{SoC_v^{t_v^a} + \eta_v P_v^0 (t-t_v^a)\Delta t, SoC_v^f}  \}
\end{equation}

\subsection{Receding Horizon Procedure}
In this section, the receding horizon approach aiming to minimize the daily peak power consumed for EV charging at the main train station is presented. Specifically, the optimization problem is solved every time the required charging power \eqref{eq:powerneed} surpasses the peak power consumption that occurred in the considered day till
the present time $t$. The optimal EV charging policy for $v \in \Omega_{EV}^t$ aiming to derive the optimal charging schedule to minimize the peak power within the moving optimization horizon $k=t, t+1, ..., T^t$ is formulated as follows:
\begin{mini!}|l|
{\text{\scriptsize $P_v^t, \lambda_p$}}{\lambda_p - \sum_{v \in \Omega_{EV}^t} \alpha_v^t  P_v^t}{\label{mod:optevpolicy11}}{}
\addConstraint{0 \leq P_v^k \leq \bar{P}_v,\quad}{k = t, ..., T^t-1}\label{mod:optevpolicy1}
\addConstraint{SoC_v^{k+1}=SoC_v^{k}+ \eta_v P_v^{k}\Delta t,\quad}{k = t, ..., T^t-1}\label{mod:optevpolicy2} 
    \addConstraint{\theta_v^k\leq SoC_v^k \leq SoC_v^f,\quad}{k = t+1, ..., T^t}\label{mod:optevpolicy4}
\addConstraint{\sum_{v \in \Omega_{EV}^t} P_v^t  \geq  \sum_{v \in\Omega_{EV}^t} P_v^k, \quad}{k = t+1, ..., T^t-1}\label{mod:optevpolicy6}\addConstraint{\hat{\lambda} \leq \sum_{v \in \Omega_{EV}^t} P_v^t  \leq \lambda_p}{  
                }\label{mod:optevpolicy5}
\addConstraint{\sum_{v \in \Omega_{EV}^t} P_v^t + P_D^t  \leq P_{max}}{ }\label{mod:optevpolicy7}
\end{mini!}
where $\lambda_p$ is the peak power for time step $t$, $P_v^t$ the charging power for vehicle $v$, $\alpha_v^t$ the weight prioritizing the charging of vehicle $v$, and $\bar{P}_v$ the maximum charging power of vehicle $v$. $SoC_v^t$ is the state of charge of vehicle $v$, $\eta_v$ the charging efficiency, $\theta_v^t$ the customer satisfaction threshold, $\hat{\lambda}$ the peak power up to time $t$, and $P_{max}$ the maximum power limit for the total railway and EV charging load at the substation of the train station.

Specifically, \eqref{mod:optevpolicy1} restricts the charging power to the allowed rate,  \eqref{mod:optevpolicy2} models the dynamics of the EV charged energy, \eqref{mod:optevpolicy4} guarantees customer satisfaction by the time of departure, \eqref{mod:optevpolicy6} determines the peak over the optimization horizon, \eqref{mod:optevpolicy5} limits the consumed power for EV charging between the current and predicted peaks, and \eqref{mod:optevpolicy7} ensures that overloading is avoided, i.e., the combined charging and railway demand is below a maximum power threshold. Thus, (219e) and (219f) guarantee that the optimal value of $\lambda_p$ represents the minimum peak power within
the time horizon and it is attained at the first time step $t$ of the moving horizon. 

\color{geo}
\section{Integrating Optimal EV Charging in Railway Energy Management}
\label{section:algo}
\subsection{Proposed EMS Algorithm}
According to the formulations in the previous section, optimal EV charging may be further exploited in the design of EMS algorithms of electric railway systems. An illustration of the structure of the proposed EMS algorithm incorporating optimal EV charging is shown in Fig. 1 which can be summarized as follows: \textbf{Steps 2-5} are for minimizing the peak power consumed by EV charging of plugged-in EVs at the railway station over the receding horizon, while \textbf{Steps 6} and \textbf{8} are for applying the optimal EV policy and solving the upper level EMS. Normally, charging is scheduled according to maximum charging rates or until the energy requirements are satisfied (Step 5a). However, once a new power peak for the EV charging requirements is observed, the optimized EV charging policy is activated (Step 5b) to derive the optimized charging schedule. 
\color{black}

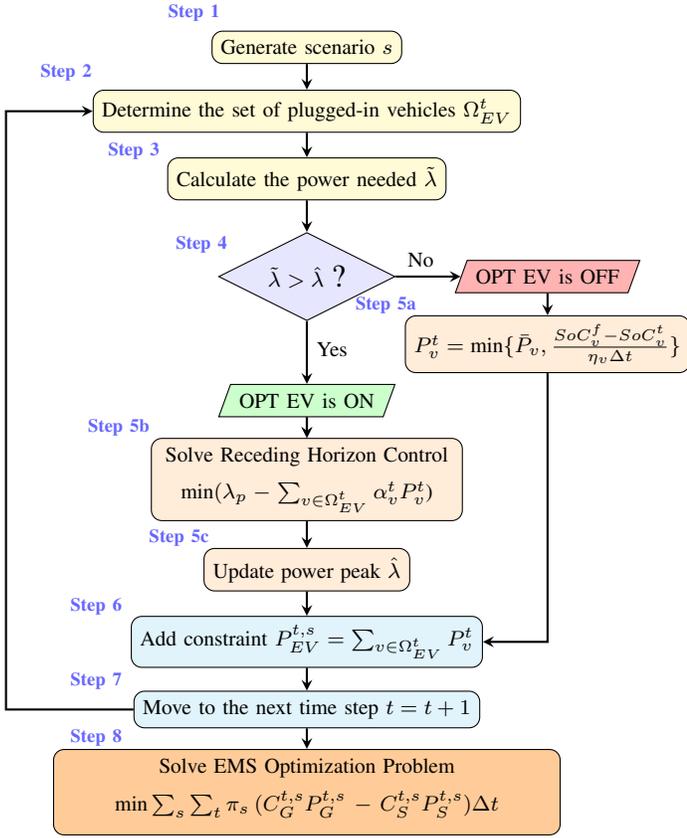
\begin{figure}[!tb]
\begin{center}
\begin{tikzpicture}[node distance=0.9cm]
[every label/.append style={text=red, font=\tiny}]
\tikzstyle{every node}=[font=\footnotesize]
\tikzstyle{every label}=[text=blue!60, font=\scriptsize]
\node (scenarios) [process, label={[xshift=-1.5cm]\textbf{Step 1}}, yshift=0.85cm, fill=yellow!20] {Generate scenario $s$};
\node (setomega) [process, label={[xshift=-3.2cm]\textbf{Step 2}}, xshift=0cm, fill=yellow!20] {Determine the set of plugged-in vehicles $\Omega_{EV}^{t}$ };
\node (powerneed) [process, label={[xshift=-2.3cm, yshift=-0.15cm]\textbf{Step 3}}, below of=setomega, xshift=0cm, fill=yellow!20] {Calculate the power needed $\tilde{\lambda}$ };
\node (check_stand) [decision, label={[xshift=-1.4cm, yshift=-0.4cm]\textbf{Step 4}}, below of =powerneed, yshift=-0.4cm, aspect=2] {$\tilde{\lambda}>\hat{\lambda}$ {\large ?}};
\node (EV_OFF) [io, right of =check_stand, xshift=2.3cm] {OPT EV is OFF};
\node (EV_OFF_ev) [process, label={[xshift=-2.15cm, yshift=-0.09cm]\textbf{Step 5a}}, below of =EV_OFF, xshift=-0cm] {${P_v^t}=\text{min} \{ \bar{P}_v, \frac{SoC_v^f-SoC_v^t}{ \eta_v \Delta t} \} $};
\node (EV_ON) [io, below of =check_stand, yshift = -0.75cm, fill=green!20] {OPT EV is ON}; 
\node (find_optev) [process, label={[xshift=-2.5cm, yshift=-0.09cm]\textbf{Step 5b}}, below of=EV_ON, text width=3.9cm, yshift=-0.15cm, xshift=-0cm] {Solve Receding Horizon Control \\ \vspace{5pt}${\text{min}} (\lambda_p - \sum_{v \in \Omega_{EV}^t} \alpha_v^t  P_v^t$) };
\node (ESS_param) [process, label={[xshift=-1.7cm, yshift=-0.09cm]\textbf{Step 5c}}, below of = find_optev, text width=2.5cm, yshift = -0.3cm] {Update power peak $\hat{\lambda}$};
\node (xi) [process,  below of = ESS_param, label={[xshift=-2.8cm, yshift=-0.1cm]\textbf{Step 6}}, yshift=-0.06cm, fill = cyan!10] {Add constraint $P_{EV}^{t,s}=  \sum_{v \in\Omega_{EV}^t} {P_v^t}$};
\node (state_matrix) [process, label={[xshift=-2.8cm, yshift=-0.1cm]\textbf{Step 7}}, below of=xi, yshift=-0.0cm, fill = cyan!10] {Move to the next time step $t=t+1$}; 
\node (tplus1) [process, text width=4cm, label={[xshift=-2.8cm, yshift=-0.08cm]\textbf{Step 8}}, below of=state_matrix, text width=6.5cm, yshift=-0.2cm, fill = orange!40] {Solve EMS Optimization Problem\\ \vspace{5pt} $ \underset {} 
 {\text{min}}  \sum_s \sum _t \pi_s \hspace{0.025in} (C_{G}^{t,s}P_G^{t,s}-C_{S}^{t,s}P_S^{t,s})\Delta t$};
 \draw [arrow] (scenarios) -- (setomega);
\draw [arrow] (setomega) -- (powerneed);
\draw [arrow] (powerneed) -- (check_stand);
\draw [arrow] (check_stand) -- node[anchor=west, xshift=0cm, yshift=0.05cm] {Yes} (EV_ON);
\draw [arrow] (EV_ON) -- (find_optev);
\draw [arrow] (EV_ON) -- (find_optev);
\draw [arrow] (find_optev) -- (ESS_param);
\draw [arrow] (ESS_param) -- (xi);
\draw [arrow] (xi) -- (state_matrix);
\draw [arrow] (state_matrix) -- (tplus1);
\draw [arrow] (check_stand) -- node[anchor=south, xshift=-0.1cm, yshift=0cm] {No} (EV_OFF);
\draw [arrow] (EV_OFF) --  (EV_OFF_ev);
\draw [arrow] (EV_OFF_ev) |- (xi);
\draw [arrow] (state_matrix) -- +(-4,0) |- (setomega);
\end{tikzpicture}
\caption{Flowchart of the proposed EMS Algorithm with optimal EV charging.}
\label{chart}
\end{center}
\label{fig:flowchart}
\vspace{-10pt}
\end{figure}
\color{geo}
\subsection{Remarks}
\begin{itemize}
\item It is worth noting that the receding horizon for the minimization of the peak EV charging is selected based on the maximum time to fulfill the charging needs of all plugged-in EVs as follows \cite{Casini21}: 
\begin{equation}
\label{eq:opthorizon}
T^t = \text{max} \{  t_v^f : v \in \Omega_{EV}^t  \}
\end{equation}
\item The proposed algorithm guarantees that EVs leave the parking lot satisfied, i.e., they have been charged with at least the average charging power as enforced by \eqref{mod:optevpolicy4}. However, customer dissatisfaction may be allowed up to some point by considering chance constraints, as proposed in \cite{Casinichance}.
\item The proposed approach is flexible regarding the number or type of EVs considered. Hence, various types of EVs, such as private electric cars of customers choosing the \qq{park and rail} option as well as electric buses arriving a few minutes before departure at the railway station may be included. 
\item The cost of RBE is not considered in the objective function of the EMS as RBE is typically wasted through thermal resistance.
In addition, investment costs are outside the scope of this work.
\end{itemize}
\color{black}

\newcommand{\MATLAB}{\textsc{Matlab}\xspace}

\color{geo}
\section{Numerical Study}
A comprehensive case study is presented in this section to demonstrate the performance of the proposed algorithm described in Section \ref{section:algo}. Firstly, the outcomes of the EV charging policy based on the receding horizon control are presented. Afterwards, the EMS model integrating optimal EV charging is simulated and the results are provided.

\subsection{Case Study Set-up}
The railway station analyzed corresponds to an actual station located in Chur, Switzerland. It is assumed that solar generation is available at the train station, whereas ESS is installed at the supply substation of the train station. To create a realistic set-up, data sets corresponding to Chur's daily train demand and RB profiles are provided by the Swiss Federal Railways and used in the simulations. Realistic electricity price signals from the day-ahead market in Switzerland are also used \cite{entsoe}, where buying price and selling price are assumed as varying over time but equal.

To calculate EV charging requirements, both private electric cars of customers choosing the \qq{park-and-rail} option as well as public electric buses at the bus stops closest to the train station are considered. The charging facilities at the train station are assumed to open to EV arrivals daily from 6:00 to 22:00, whereas departure may happen either within or outside opening hours. For private EV cars, arrival times follow an exponential distribution with a rate of 4, i.e., an average of 4 cars is expected per hour, while departure times are chosen based on a triangular distribution within 2 hours from their fulfillment time. It is also assumed that private EVs have a nominal charging power of 11 kW and a maximum charging power of 22 kW and their required charging energy follows a uniform distribution in the interval [0, 50] kWh. For public electric buses, departure times are according to the public schedule as posted on \cite{postauto}, whereas arrival times are chosen based on a triangular distribution between 10 and 30 minutes from their departure. Electric buses have typically greater charging requirements and limits. Hence, their required charging energy follows a uniform distribution in the interval [0, 300] kWh. The nominal charging power and the maximum charging power for electric buses are set as 300 kW. 

To create the daily trajectories for the solar generated power, actual daily solar radiation data provided by the Swiss Federal Railways are used in \eqref{eq:solar_p_val}. The assumed installed solar capacity at the train station is $P_r = 1000$ kW, which corresponds to 20$\%$ of the peak train demand observed in the data. The rest of the parameters are set as $r_c=150$ $\text{W/m}^2$ and $r_{std}=1000$ $\text{W/m}^\text{2}.$

The ESS model is assumed to have a capacity of $1000$ kWh. Besides, charging and discharging rates are set as $1000$ kW/min, the self-discharge coefficient is $\epsilon_{B-}=0$ and the charging and discharging efficiencies are $\eta_{B+}=\eta_{B-}=0.95$. The initial energy level for the ESS is set as $50\%$ of the ESS capacity. The minimum energy limit for the ESS is taken as $10\%$ of the ESS capacity.

The proposed EMS algorithm integrating optimal EV charging is implemented in \textsc{Matlab} $^\copyright$. The optimization problems in \textbf{Step 5b} and \textbf{Step 8} are solved with the commercially available linear programming solver Gurobi \cite{gurobi}. To account for issues arising due to computational burden, a sampling time of 10 minutes is selected.

\subsection{Validation of the Proposed Algorithm}
\label{casestudy1}

To assess the performance of the proposed railway EMS integrating optimal EV charging, numerical simulations considering 150 scenarios of equal probability $\pi_s$ corresponding to different days, solar radiation, and train demand profiles over the year 2021 are performed. 

\begin{figure}[!t]
\centering
\includegraphics[width=2.75in ,keepaspectratio=true,angle=0]{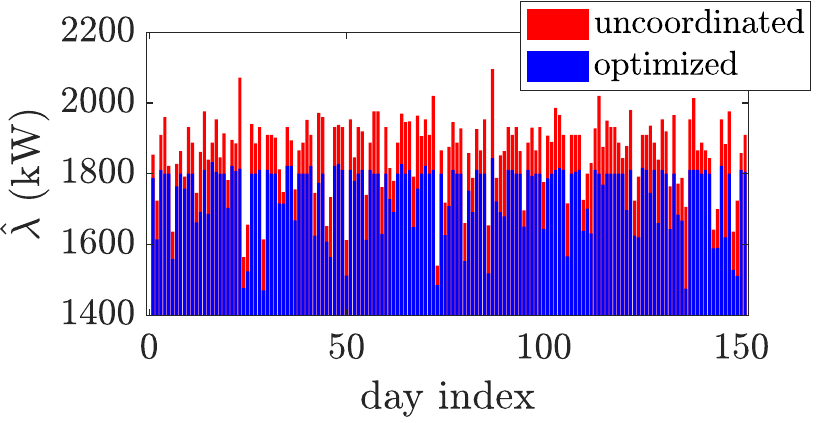}
\caption{Optimized EV charging power peaks for 150 days/scenarios.}
\label{fig:peaks}
\end{figure}

Fig. \ref{fig:peaks} depicts the daily peak powers as achieved following the optimized EV charging policy for the 150 different scenarios. Particularly, the results of the optimized EV charging policy presented in Section III are compared to the ones of the uncoordinated charging policy, where each vehicle is simply charged with a constant maximum charging power until the satisfaction of its charging requirements or its departure. It can be observed that the optimized charging schedule that relies on receding horizon control leads to peak powers that are always smaller than the ones obtained by uncoordinated charging. Thanks to the optimized schedule, peak power savings of up to 14.01$\%$ can be achieved, which prevent overloading and could further have a positive impact on system operation when operating close to or beyond the technical constraints, e.g., line limits.

Next, optimized EV charging schedules for each scenario are added as constraints (\textbf{Step 6}) and railway EMS is performed (\textbf{Step 8}). To highlight the effectiveness of the proposed algorithm, two different cases are considered. \textit{Case 1} corresponds to the proposed railway EMS with ESS, RBE,
and solar generation included to meet the combined railway demand and optimized EV charging demand. \textit{Case 2} represents the base case, as no ESS, RBE or solar generation are included; only grid power is exploited to serve the railway and EV charging demand.

\begin{figure}[!tb]
\centering
\subfloat[$\text{Train demand}$]{\includegraphics[width=1.695in]{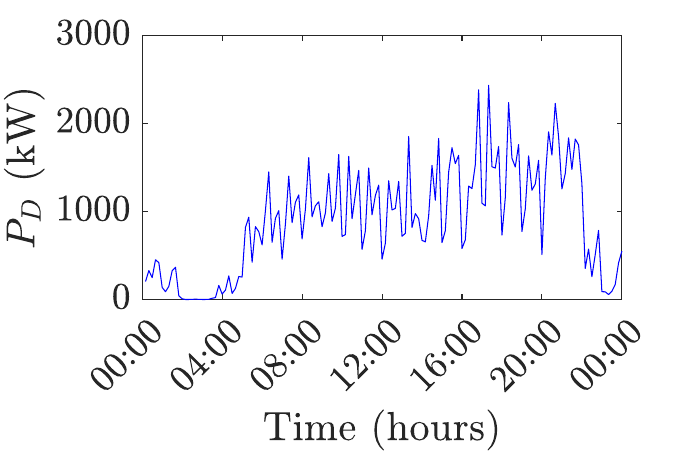}
\label{demand1}}
\hfil
\subfloat[RB power]{\includegraphics[width=1.695in]{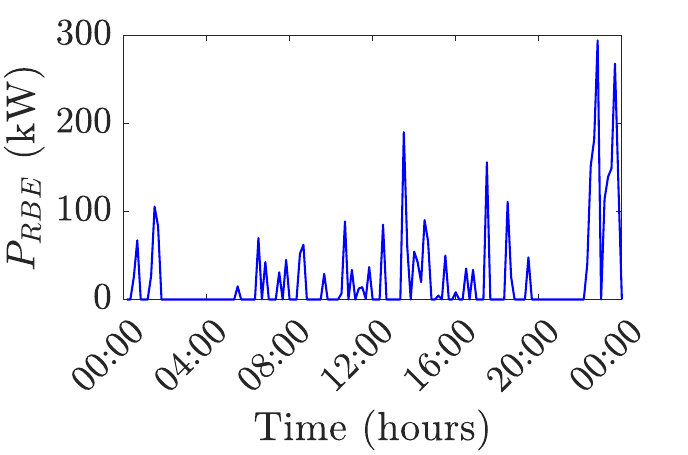}
\label{demand2}}
\caption{The railway consumption and RB profile at Chur station during the 82-th day/scenario.} \label{demands} 
\subfloat[Solar generation]{\includegraphics[width=1.695in]{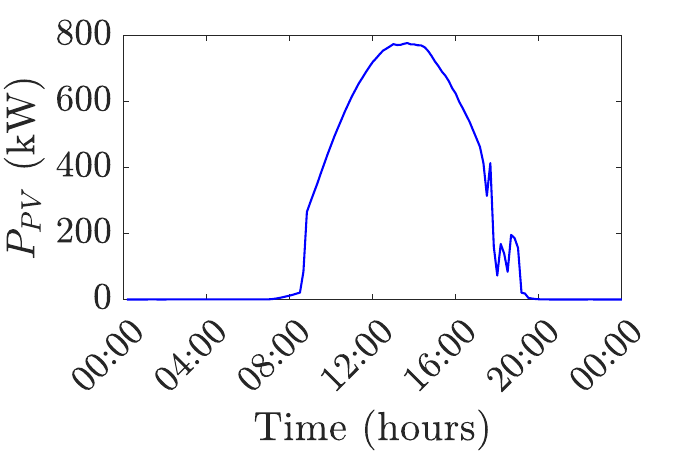}
\label{solar1}}
\hfil
\subfloat[Day-ahead price]{\includegraphics[width=1.695in]{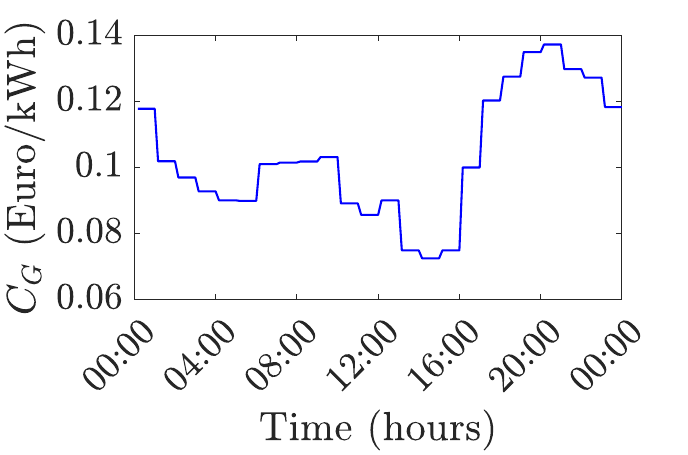}
\label{price2}}
\caption{The available solar power and day-ahead electricity price during the 82-th day/scenario.}
\label{rbeprice}
\subfloat[EV demand]{\includegraphics[width=1.695in]{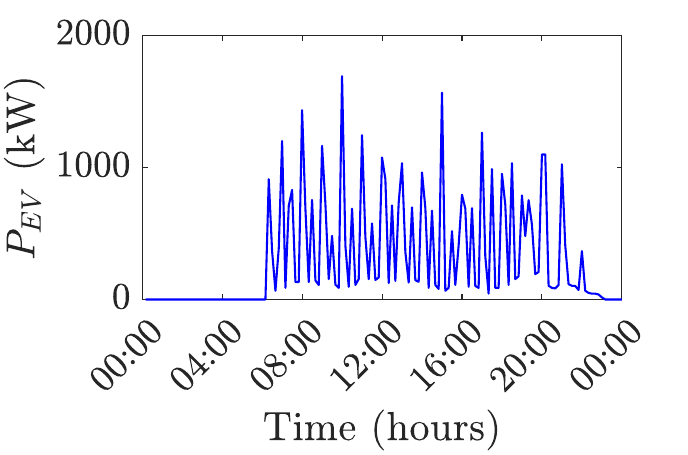}
\label{soe1}}
\hfil
\subfloat[ESS energy level]{\includegraphics[width=1.695in]{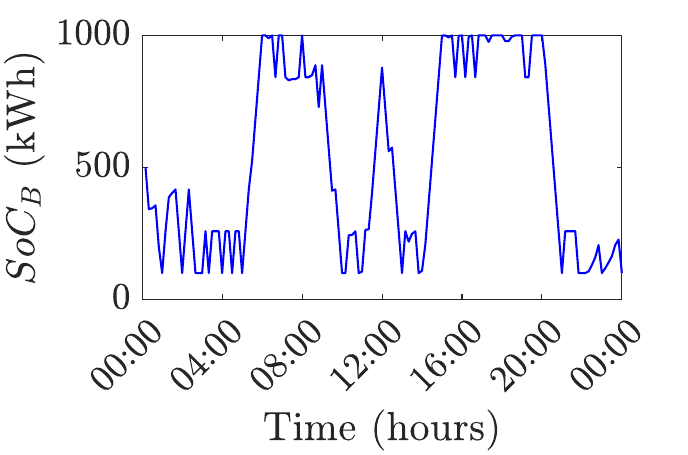}
\label{ssoe2}}
\caption{The optimized EV charging schedule and the energy level of the ESS during the 82-th  day/scenario.}
\label{soe}
\end{figure}

The results in terms of the average daily operating cost considering 150 scenarios for the two different cases are given in Table \ref{table1}. It can be seen that the proposed EMS, leveraging on ESS, RBE, and renewable generation, leads to cost savings of 17.29$\%$ for the considered scenarios compared to the base case. To further illustrate the performance of the algorithm, Figs. \ref{demands}--\ref{soe} show the daily train demand, RB availability, solar generation, electricity price, optimized EV charging schedule, and ESS energy level during the 82-th day of the simulation. Thanks to the integrated EV charging optimization, the combined railway and EV requirements are well-scheduled and lower than the line capacity. It can be verified that the implemented EMS achieves an efficient system operation with respect to variations in input data, such as RB power, solar generation, and varying prices. Indeed, looking at the ESS behavior in Fig. \ref{ssoe2}, it can be observed that ESS stores energy when RB power is available and prices are generally lower (e.g., 4:00-6:00, 10:00-12:00, 14:00-16:00) whereas it discharges when prices are higher (e.g., 20:00-22:00). In addition, ESS optimally functions to serve the combined railway and optimized EV charging loads in coordination with PV generation when there is high PV availability (e.g., 08:00-10:00, 12:00-14:00). It is worth noting that RB power may not be always stored but it can also be sold back to the grid during high price hours (e.g, 22:00-00:00). Hence, the proposed method can efficiently coordinate the different components while ensuring that operating costs are minimized and technical constraints are respected.

\color{black}

\begin{table}[!tb]
\centering
  \caption{Results - 150 Scenarios}\label{table1}
   \setlength{\tabcolsep}{2pt}
  \begin{tabular}{|c|c|c|c|c|}
\hhline{|-|-|-|-|-|}
\textbf{Case}& \textbf{ESS} & \textbf{PV} & \textbf{Total Operating Costs (€)} & \textbf{Cost Savings} (\%)
\\
\hline
1&\checkmark&\checkmark&3316.09&17.29
\\

\hline
2&-&-&4009.09&-\\
\hhline{|-|-|-|-|-|}
  \end{tabular}
  
\end{table}

\section{Conclusion}
\color{geo}This paper proposes a novel EMS algorithm for EV charging in electric railway stations with ESS and renewable generation. As opposed to previous works, the proposed method integrates an EV charging policy to optimize charging schedule and avoid potential grid issues, such as overloading. Specifically, receding horizon control is leveraged to minimize the daily peak power spent on EV charging, achieving peak power savings of up to 14.01$\%$ for the considered scenarios compared to uncoordinated charging. The optimized EV schedule is then incorporated into the railway EMS with ESS, RBE, and PV generation to minimize the daily operating costs.  Indeed, economic benefits of 17.29$\%$ compared to the base case were observed. This work represents the first attempt to account for the combined railway and EV demand while considering peak load capacity in railway EMS. Future efforts will be devoted to considering EV charging flexibility to allow for a tolerance on charging service and customer satisfaction.
\color{black}

\color{geo}
\section*{Acknowledgments}
The authors would like to thank Robert Strietzel and the Swiss Federal Railways for providing the railway consumption and solar radiation data and for interesting discussions.
\color{black}



 

  \bibliography{MyReferences}

\begin{thebibliography}{10}
\providecommand{\url}[1]{#1}
\csname url@samestyle\endcsname
\providecommand{\newblock}{\relax}
\providecommand{\bibinfo}[2]{#2}
\providecommand{\BIBentrySTDinterwordspacing}{\spaceskip=0pt\relax}
\providecommand{\BIBentryALTinterwordstretchfactor}{4}
\providecommand{\BIBentryALTinterwordspacing}{\spaceskip=\fontdimen2\font plus
\BIBentryALTinterwordstretchfactor\fontdimen3\font minus
  \fontdimen4\font\relax}
\providecommand{\BIBforeignlanguage}[2]{{%
\expandafter\ifx\csname l@#1\endcsname\relax
\typeout{** WARNING: IEEEtran.bst: No hyphenation pattern has been}%
\typeout{** loaded for the language `#1'. Using the pattern for}%
\typeout{** the default language instead.}%
\else
\language=\csname l@#1\endcsname
\fi
#2}}
\providecommand{\BIBdecl}{\relax}
\BIBdecl

\bibitem{7587629}
J.~C. Hernandez and F.~S. Sutil, ``Electric vehicle charging stations feeded by
  renewable: {PV} and train regenerative braking,'' \emph{IEEE Latin America
  Transactions}, vol.~14, no.~7, pp. 3262--3269, 2016.

\bibitem{9024071}
M.~Brenna, F.~Foiadelli, and H.~J. Kaleybar, ``The evolution of railway power
  supply systems toward smart microgrids: The concept of the energy hub and
  integration of distributed energy resources,'' \emph{IEEE Electrification
  Magazine}, vol.~8, no.~1, pp. 12--23, 2020.

\bibitem{7482762}
J.~A. Aguado, A.~J. Sánchez~Racero, and S.~de~la Torre, ``Optimal operation of
  electric railways with renewable energy and electric storage systems,''
  \emph{IEEE Transactions on Smart Grid}, vol.~9, no.~2, pp. 993--1001, 2018.

\bibitem{9782535}
Y.~Ge, H.~Hu, J.~Chen, K.~Wang, and Z.~He, ``Combined active and reactive power
  flow control strategy for flexible railway traction substation integrated
  with {ESS} and {PV},'' \emph{IEEE Transactions on Sustainable Energy},
  vol.~13, no.~4, pp. 1969--1981, 2022.

\bibitem{Sengor18}
I.~Şengör, H.~C. Kılıçkıran, H.~Akdemir, B.~Kekezo\u{g}lu, O.~Erdinç,
  and J.~P.~S. Catalão, ``Energy management of a smart railway station
  considering regenerative braking and stochastic behaviour of {ESS} and {PV}
  generation,'' \emph{IEEE Transactions on Sustainable Energy}, vol.~9, no.~3,
  pp. 1041--1050, 2018.

\bibitem{Sarabi19}
S.~Sarabi, A.~Davigny, Y.~Riffonneau, and B.~Robyns, ``{V2G} electric vehicle
  charging scheduling for railway station parking lots based on binary linear
  programming,'' in \emph{2016 IEEE International Energy Conference
  (ENERGYCON)}, Leuven, Belgium, 2016.

\bibitem{Fernandez17}
A.~Fernandez-Rodriguez, A.~Fernandez-Cardador, A.~De~Santiago-Laporte,
  C.~Rodriguez-Sanchez, A.~P. Cucala, A.~J. Lopez-Lopez, and R.~R. Pecharroman,
  ``Charging electric vehicles using regenerated energy from urban railways,''
  in \emph{2017 IEEE Vehicle Power and Propulsion Conference (VPPC)}, Belfort,
  France, 2017.

\bibitem{9745034}
A.~Çiçek, I.~Şengör, S.~Güner, F.~Karakuş, A.~K. Erenoğlu, O.~Erdinç,
  M.~Shafie-Khah, and J.~P.~S. Catalão, ``Integrated rail system and {EV}
  parking lot operation with regenerative braking energy, energy storage system
  and {PV} availability,'' \emph{IEEE Transactions on Smart Grid}, vol.~13,
  no.~4, pp. 3049--3058, 2022.

\bibitem{WECC}
\BIBentryALTinterwordspacing
{WECC Renewable Energy Modeling Task Force}, ``{WECC guide for representation
  of photovoltaic systems in large-scale load flow simulations},'' 2010.
  [Online]. Available: \url{https://www.wecc.org}
\BIBentrySTDinterwordspacing

\bibitem{Pierrouu19}
G.~Pierrou and X.~Wang, ``The effect of the uncertainty of load and renewable
  generation on the dynamic voltage stability margin,'' in \emph{2019 IEEE PES
  Innovative Smart Grid Technologies Europe (ISGT-Europe)}, Bucharest, Romania,
  2019.

\bibitem{8260299}
I.~Sengor, H.~C. Kılıçkıran, H.~Akdemir, B.~Kekezoglu, O.~Erdinç, and
  J.~P.~S. Catalão, ``Smart railway station energy management considering
  regenerative braking and {ESS},'' in \emph{2017 IEEE PES Innovative Smart
  Grid Technologies Conference Europe (ISGT-Europe)}, Turin, Italy, 2017.

\bibitem{Casini21}
M.~Casini, A.~Vicino, and G.~G. Zanvettor, ``A receding horizon approach to
  peak power minimization for {EV} charging stations in the presence of
  uncertainty,'' \emph{International Journal of Electrical Power and Energy
  Systems}, vol.~9, no.~1, 2021.

\bibitem{Casinichance}
\vspace{0mm}M. Casini, A.~Vicino, and G.~G. Zanvettor, ``A chance constraint
  approach to peak mitigation in electric vehicle charging stations,''
  \emph{Automatica}, vol. 131, no.~1, 2021.

\bibitem{entsoe}
\BIBentryALTinterwordspacing
{ENTSO-E Transparency Platform}, ``{Day-ahead prices},'' 2022. [Online].
  Available: \url{https://transparency.entsoe.eu/}
\BIBentrySTDinterwordspacing

\bibitem{postauto}
\BIBentryALTinterwordspacing
{Chur, Postautostation Departures}, ``{Chur, Postautostation},'' 2022.
  [Online]. Available:
  \url{https://timetable.search.ch/Chur,Postautostation?time_type=depart}
\BIBentrySTDinterwordspacing

\bibitem{gurobi}
\BIBentryALTinterwordspacing
{Gurobi Optimization, LLC}, ``{Gurobi optimizer reference manual},'' 2022.
  [Online]. Available: \url{https://www.gurobi.com}
\BIBentrySTDinterwordspacing

\end{thebibliography}
\bibliographystyle{IEEEtran}

\vfill

\end{document}